\documentclass[preprint]{revtex4}
\usepackage{epsfig}

\begin{document}

\title{The three-body recombination of a condensed Bose gas near a Feshbach resonance}
\author{Yingyi Zhang, Lan Yin}
\email{yinlan@pku.edu.cn} \affiliation{School of Physics, Peking
University, Beijing 100871, China}
\date{July 6, 2005}

\begin{abstract}

In this paper, we study the three-body recombination rate of a
homogeneous dilute Bose gas with a Feshbach resonance at zero
temperature. The ground state and excitations of this system are
obtained. The three-body recombination in the ground state is due to
the break-up of an atom pair in the quantum depletion and the
formation of a molecule by an atom from the broken pair and an atom
from the condensate.  The rate of this process is in good agreement
with the experiment on $^{23}$Na in a wide range of magnetic fields.

\end{abstract}

\maketitle
\section{Introduction}
Two-body and three-body recombinations are the main reasons for
particle loss in cold Bose gases.  Usually the two-body
recombination is dominant in the dilute limit and the three-body
recombination becomes more important when either the particle
density or the interaction strength increases.  In systems with
Feshbach resonances, near a resonance the scattering length becomes
very large.  It was found in the experiment on $^{23}$Na
\cite{Stenger} that the particle-loss rate increases enormously when
close to a Feshbach resonance.  The three-body recombination rate
obtained from the experimental data is much higher than an earlier
theoretical estimate for a simple dilute Bose gas \cite{Fedichev}.

The large particle-loss rate in $^{23}$Na was explained by one of us
in terms of the many-body effect due to the special ground-state
structure of a dilute condensed Bose gas with a Feshbach resonance
\cite{Yin}. The ground state contains both the condensate and the
quantum depletion. One pair of atoms from the quantum depletion can
break up.  One atom from the broken pair and an atom from the
condensate can form a molecule. Therefore in the final state, these
three atoms become a molecule and an atom with opposite momenta. The
rate of this process was found to be in good agreement with the
experimental data close to the resonance \cite{Yin}.

There have been considerable progress in the study of the three-body
recombination in few-body systems
\cite{Hammer,Braaten1,Braaten2,Esry,Petrov}. (See Ref. \cite{Hammer}
for a complete review on this subject). Particularly, in Ref.
\cite{Petrov}, the large particle-loss rate in $^{23}$Na was
explained in terms of the few-body physics. One important prediction
from these studies is that the three-body recombination rate
displays periodic behavior due to Efimov states, which has not been
observed in experiments so far.

However both theoretical results from Ref. \cite{Yin} and Ref.
\cite{Petrov} can only fit the experimental data of the
particle-loss rate \cite{Stenger} very close to the resonance.
Theoretical explanation is still needed for the experimental data
away from the resonance. In this paper, we adopt an approach which
is valid not only close to the resonance, but also quite far away
from the resonance, and study a dilute homogeneous system at zero
temperature.

\section{The Two-channel model}
The starting point of our approach is the two-channel model
\cite{E. Timmermans} for systems with a Feshbach resonance,
\begin{equation}\label{H}
H=\frac{\hbar^2}{2m}\nabla\psi^{\dagger}\cdot\nabla\psi+
\frac{\hbar^2}{4m}\nabla\phi^{\dagger}\cdot\nabla\phi+
\frac{g_0}{2}\psi^{\dagger{2}}\psi^2-
\gamma(\phi^{\dagger}\psi^2+h.c.)+h\phi^{\dagger}\phi,
\end{equation}
where $m$ is the atom mass, $\phi$ is the molecular-field operator,
and $\psi$ is the atomic-field operator.  The magnetic detuning
energy is given by $h$. The coupling constant of atoms in absence of
the resonance is given by $g_0=4\pi\hbar^2a_0/m$, where $a_0$ is the
asymptotic scattering length far away from the resonance.  The
coupling constant between atoms and molecules is given by $\gamma$,
from which we can define a length scale $a_\gamma \equiv 2\pi
\hbar^4/(m^2 \gamma^2)$, corresponding to the effective range in the
two-atom scattering problem \cite{Ho}.

In the two-channel model, there are conversions between atoms and
molecules. It is usually convenient to work in the grand canonical
ensemble with the grand thermodynamical potential given by
\begin{equation}
F=H-\mu(\psi^{\dagger}\psi+2\phi^{\dagger}\phi),
\end{equation}
where $\mu$ is the chemical potential.

\section{The ground state structure}
In the dilute region, at zero temperature, both the atoms and the
molecules condense.  The condensed atoms and molecules have phase
coherence with each other, and together form the condensate.  The
condensate provides the dominant contribution to the grand
potential, which is given by
\begin{equation}
F_0=\frac{g_0}{2}|\psi_0|^4-\gamma(\phi_0^{*}\psi_0^2+c.c.)
+h|\phi_0|^2-\mu(|\psi_0|^2+2|\phi_0|^2),
\end{equation}
where $\psi_0$ and $\phi_0$ are the expectation values of the
atomic and molecular fields. The value of $\psi_0$ and $\phi_0$
are determined from the saddle-point equations
\begin{equation}
\frac{\partial
F_0}{\partial\psi_0}=(g_0|\psi_0|^2-\mu)\psi_0^*-2\gamma\phi_0^*
\psi_0=0,
\end{equation}
\begin{equation}
\frac{\partial F_0}{\partial
\phi_0}=(h-2\mu)\phi_0^*-\gamma\psi_0^{*2}=0.
\end{equation}
In the general solution of these saddle-point equations, the
molecular field is always in phase or out of phase with the square
of the atomic field.  For convenience, we choose the expectation
value of the atomic field to be positive, $\psi_0>0$.

In the region with the repulsive interaction, the detuning energy
is negative, $h<0$, and the chemical potential is positive,
$\mu>0$.   The non-trivial solution of the saddle-point equations
are given by
\begin{equation}
\psi_0^2=\frac{\mu}{g_{\rm eff}},
\end{equation}
\begin{equation}
\phi_0=\frac{\gamma\psi_0^2}{h-2\mu},
\end{equation}
where $g_{\rm eff}=g_0-2\gamma^2/(h-2\mu)$ is the effective
coupling constant between atoms.  The scattering length is
proportional to the effective coupling constant,
\begin{equation}\label{8}
a=mg_{\rm eff}/(4\pi\hbar^2)=a_0(1-\frac{\Delta}{h-2\mu}),
\end{equation}
where the width of the resonance is given by $\Delta=2\gamma^2/g_0$.
In the ground state of a dilute Bose gas, most of the particles are
condensed atoms \cite{Yin}, $\psi_0^2 \approx n$, where $n$ is the
total particle density.  The chemical potential $\mu$ is much
smaller than other energy quantities which do not have density
dependence.

However as in the traditional theory of a dilute Bose gas, the
condensate is not the complete picture of the ground state.  The
ground state also contains the quantum depletion which is made of
pairs of atoms with opposite momenta.  To describe the quantum
depletion, we need to look at the gaussian fluctuation around the
condensate. In the long-wavelength and low-energy limit, the
gaussian fluctuation in the grand potential is given by
\begin{equation}\label{quad}
\delta F_2=\sum_{\bf
k}[(\epsilon_k-\mu+2g_0\psi_0^2-{2\gamma^2\psi_0^2 \over
h-2\mu})\psi_{\bf k}^\dagger \psi_{\bf k}+({g_0 \psi_0^2 \over
2}-\gamma \phi_0) ({\psi_{\bf k}^\dagger} \psi_{\bf
-k}^\dagger+h.c.)],
\end{equation}
where $\epsilon_k$ is the kinetic energy of the atom,
$\epsilon_k=\hbar^2 k^2/(2m)$.  The molecular part of the
fluctuation is not included in equation (\ref{quad}) because
molecules have finite energy in the long-wavelength limit. However
through the virtual excitation of a molecule, the atoms acquire a
diagonal self-energy approximately given by $ -2\gamma^2\psi_0/(
h-2\mu)$, which is included in the r.h.s. of Eq.(\ref{quad}).

After applying Bogoliubov transformation, we obtain
\begin{equation}
\delta F_2=C+\sum_{\bf k}E_k c_{\bf k}^\dagger c_{\bf k},
\end{equation}
where the transformation is given by $c_{\bf k}=u_k \psi_{\bf
k}+v_k \psi_{-{\bf k}}^\dagger$, with
$u_k^2=[1+(\epsilon_k+\mu)/E_k]/2$ and $v_k^2=u_k^2-1$. The atoms
have a gapless phonon mode with the excitation energy given by
\begin{equation}\label{Ek}
E_k=\sqrt{\epsilon_k(\epsilon_k+2\mu)}.
\end{equation}
In the ground state $|G\rangle$, there are no phonon excitations,
$c_{\bf k}|G\rangle=0$, which is only possible if in the ground
state atoms with opposite wave-vectors ${\bf k}$ and ${-\bf k}$ are
paired up. The paired atoms form the quantum depletion with the
depletion energy $C$ given by
\begin{equation}\label{Eg}
{C \over V}={8 \over 15\pi^2}\left({m \mu\over \hbar^2}\right)^{3
\over 2} \mu,
\end{equation}
where $V$ is the total volume of the system.  As we will discuss in
the later sections, the ground state obtained here is not the state
with the lowest energy, but a metastable state.  Nonetheless, in the
contest of the traditional theory of a dilute Bose gas, we still
name this state as the ground state in this paper.

\section{Vacuum renormalization to the molecular excitations}
Beside the ground state and phonon excitations, the system also
contains molecular excitations.  From equation (\ref{H}), the bare
molecular excitation energy is simply given by the detuning energy
$h$ in the long-wavelength limit.  However, the molecule energy is
strongly renormalized by the repetitive process of molecules
turning into and forming from atoms.  In the dilute region, the
renormalization to the molecule energy due to the many-body
interaction is much smaller than vacuum renormalization because
the gas parameter is much smaller than one, $\sqrt{8\pi n a^3}\ll
1$.   In the following, we study the vacuum renormalization in
detail.

\begin{figure}
\centering \epsfig{file=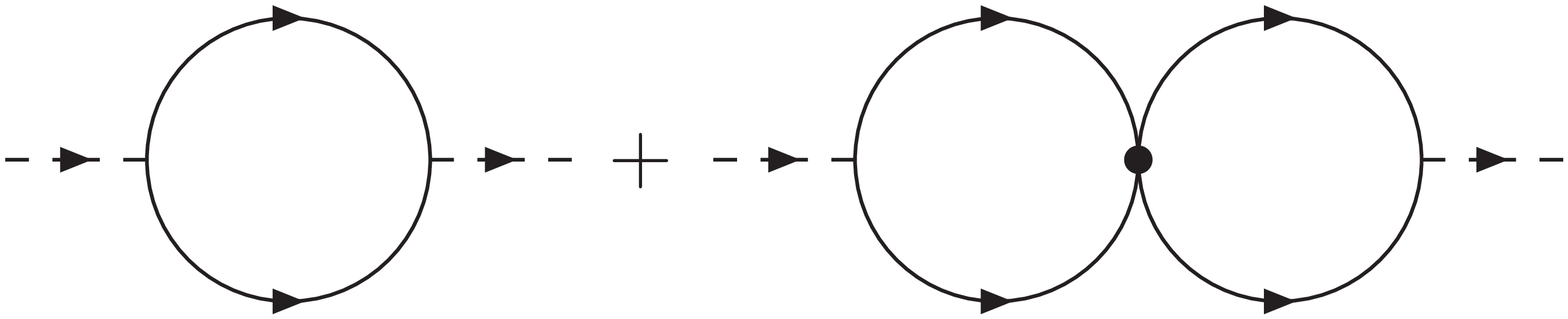, height=1.5in} \label{self-energy}
\caption{The self-energy diagrams of the molecular propagator.  The
solid lines represent the atom propagator and the dashed lines
represent the molecule propagator.  The vertex represents the
coupling constant $g_0$.} \label{mseg}
\end{figure}

In the vacuum, the propagator of the molecular field is given by
\begin{equation}
G_m({\bf q},\Omega)=\frac{1}{\Omega-\frac{\epsilon_q}{2}-h+2\mu
-\Sigma_m({\bf q},\Omega)+i\delta},
\end{equation}
where $\Sigma_m({\bf q},\Omega)$ is the self-energy due to the
vacuum renormalization.   The self-energy diagrams of the molecular
propagator are shown in Fig. \ref{mseg}.  As discussed in earlier
references \cite{Yin,Stoof,Bruun}, the self-energy of molecules is
given by
\begin{equation}\label{mse}
\Sigma_m({\bf q},\Omega)=2\gamma^{2}D({\bf q},\Omega)[1+g_0 D({\bf
q},\Omega)],
\end{equation}
where
\begin{equation}\label{D-f}
D({\bf q},\Omega) \equiv i\int\frac{d\omega}{2\pi}\int
\frac{d^{3}k}{(2\pi)^3} G_a({\bf k},\omega) G_a({\bf q}-{\bf
k},\Omega-\omega),
\end{equation}
and the propagator of the atomic field is given by
\begin{equation}
G_a({\bf k},\omega)=\frac{1}{\omega-\epsilon_k+\mu+i\delta}.
\end{equation}

In Ref. \cite{Yin}, the second term in the r.h.s of equation
(\ref{mse}) was dropped.  As a result, the calculations in Ref.
\cite{Yin} can only be applied when the system is close to the
resonance.  Here we keep both terms in the expression of the
molecular self-energy in equation (\ref{mse}).

As discussed in Ref. \cite{Yin}, the divergence in the
$D$-function in equation (\ref{D-f}) can be removed following the
standard renormalization procedure by introducing a counter term,
which yields
\begin{eqnarray}
D({\bf q},\Omega) &=& \int\frac{d^{3}k}{(2\pi)^3}
(\frac{1}{\Omega-\epsilon_k-\epsilon_{k-q}
+2\mu+i\delta}+\frac{1}{2\epsilon_k}) \nonumber \\
&=&{m^{3/2} \over 4\pi \hbar^3}
\sqrt{\frac{\epsilon_q}{2}-2\mu-\Omega}.
\end{eqnarray}
The molecular self-energy is thus given by
\begin{equation} \label{mse2}
\Sigma_m({\bf q},\Omega)=\nu \sqrt{\frac{\epsilon_q}{2}-2\mu-\Omega}
+ c(\frac{\epsilon_q}{2}-2\mu-\Omega),
\end{equation}
where the parameters $\nu$ and $c$ are given by
\begin{eqnarray}
\nu & \equiv& {m^{3/2} \gamma^2 \over 2\pi \hbar^3}=\sqrt{-h (a-a_0)
\over a_\gamma} \\ c &\equiv& {m^3 \gamma^2 g_0 \over 8\pi^2
\hbar^6}={a_0 \over a_\gamma}.
\end{eqnarray}

Although the self-energy given by equation (\ref{mse2}) looks rather
complicated, the main feature of the the molecular propagator is a
pole which is given by
\begin{equation}
\Omega_q =\frac{\epsilon_q}{2}-2\mu-{[\sqrt{\nu^2-4(c+1)h}-\nu]^2
\over 4(c+1)^2}.
\end{equation}
Close to the pole, the propagator can be approximated as
\begin{equation}
G_m({\bf q},\Omega) \approx \frac{Z}{\Omega-\Omega_q+i\delta},
\end{equation}
where the renormalization factor $Z$ is given by
\begin{eqnarray}
Z &=& \left[1+c+{\nu \over 2\sqrt{{\epsilon_q \over 2} -2 \mu-
\Omega_q}}\right]^{-1} \nonumber \\
&=& \left[1+c+{\nu (c+1) \over \sqrt{\nu^2-4(c+1)h}-\nu}
\right]^{-1}.
\end{eqnarray}
In the limit $a \gg a_0$, the molecule energy is approximately given
by the shallow bound-state energy, $\Omega_0 \approx -\hbar^2/
(a^2m)$, as given in Ref. \cite{Yin}.  In the limit that $a$ is very
close to $a_0$, the molecule energy is approximately given by the
detuning energy, $\Omega_0 \approx h$.

\section{The three-body recombination}
In the repulsive-interaction region, the molecules have lower energy
than the atoms.  Thus the system can always reduce its energy by
forming more molecules.  The ground state obtained so far is not the
state with the lowest energy, but a metastable state. The three-body
recombination is one of the main processes for the system to move
away from the metastable state.

In the metastable condensed state, the three-body recombination
takes place in the following manner. One pair of atoms with opposite
momenta in the quantum depletion break up.  One atom from the broken
pair can interact with an atom in the condensate to form a molecule,
leaving the other atom in the broken pair as an excited atom. Thus
the final state of this process is the ground state plus a molecule
and an excited atom. The probability density of the three-body
recombination process can be obtained by using Fermi's golden rule
\begin{eqnarray} \label{rate}
\Gamma &=&{2\pi \over \hbar}\int\frac{d^3 k}{(2\pi)^3} |\langle{\bf
k}| 2 \gamma\psi_0\phi_{\bf k}^{\dagger} \psi_{\bf k} |G\rangle|^2
\delta(E_k+\Omega_k) \nonumber \\
&=&\frac{2\pi}{\hbar}\int\frac{d^3 k} {(2\pi)^3}4\gamma^2 \psi_0^2
v_k^2 Z \delta(E_k+\Omega_k) \nonumber \\
&\approx& \frac{2\pi}{\hbar}\int\frac{d^3 k} {(2\pi)^3}4\gamma^2 n
{\mu^2 \over 4\epsilon_k^2} Z \delta({3\over 2}\epsilon_k+\Omega_0) \nonumber \\
&=& {16 \sqrt{3}\pi\gamma^2Za^2n^3\over\sqrt{m |\Omega_0|^3}},
\end{eqnarray}
where $|{\bf k}\rangle$ is the final state which is the ground state
plus a molecule with wave vector ${\bf k}$ and an excited atom with
wave vector $-{\bf k}$.

\begin{figure}[ht!p]
\centering
\includegraphics[width=0.6\linewidth]{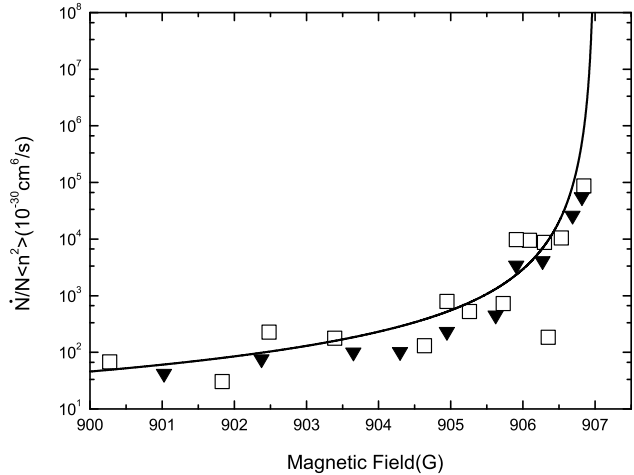}
\caption{The coefficients of the particle-loss rate for the Na
system,  $\dot{N}/N\langle n^2\rangle$ is plotted vs the magnetic
field as in Fig. (2) of Ref. \cite{Stenger}. The straight line is
the theoretical result computed from Eq.(\ref{rate}); the squares
and triangles are the experimental data taken at ramp speeds
0.13G/ms and 0.31G/ms of the magnetic field\cite{Stenger}. All the
parameters used in the calculation are taken from the experiment
\cite{Stenger}.} \label{final}
\end{figure}

In the limit that $a \gg a_0$, the three-body recombination rate
$\Gamma$ is proportional to $a^4$ \cite{Yin}, with a coefficient
about 140 times larger than that in the single channel case
\cite{Fedichev}.  Away from this limit, although there is no simple
scaling form for the rate $\Gamma$, we can compute its numerical
value from equation (\ref{rate}).  Here the theoretical value of the
particle-loss rate in the $^{23}$Na system is plotted and compared
with the experimental data \cite{Stenger} in Fig. \ref{final}. The
experimental data is the loss rate divided by the product of total
particle number $N$ and the average of the density squared. The
theoretical value is the loss rate divided by the product of $N$ and
the homogeneous density squared.  All the parameters used in the
calculation are taken from the experiment, $a_0=3.3nm$ and $\Delta=2
\mu_B G$. The theoretical result is fairly close to the experimental
data in a wide range of magnetic fields in the repulsive-interaction
region, although the theoretical derivation so far is for the
homogeneous system and the experimental system is inhomogeneous.
\\

\section{Conclusions}
We have studied a homogeneous dilute Bose gas with a Feshbach
resonance is studied at zero temperature.  In the ground state,
there are both the condensate and the quantum depletion.  There are
two types of excitations, phonon excitations and molecular
excitations.  The three-body recombination in the ground state is
due to the formation of a molecule from an atom in the condensate
and an atom from the quantum depletion. The excellent agreement
between our calculation and the experimental results in a wide range
of magnetic fields leads us to conclude that the many-body effect
due to the ground-state structure plays a major role in the
three-body recombination process.

An interesting question is whether or not such theoretical
description is still valid when the system moves further away from
the resonance.  When the system is very far from the resonance, the
two-channel description should probably be replaced by a
single-channel model and the three-body recombination rate is likely
to recover the single-channel form.  One possible criterion is the
comparison between the molecule energy in the two-channel model and
the bound-state energy in the single-channel model.  If the molecule
energy in the closed channel is much deeper than the bound-state
energy in the open channel, the single-channel description is
probably better.

We would like to thank T.-L. Ho, H.-W. Hammer and E. Braaten for
helpful discussions. This work is supported by NSFC under Grant No.
90303008, by Key Project of Chinese Ministry of Education, and by
SRF for ROCS, SEM.

\end{document}